\documentclass[pra,twocolumn,showpacs]{revtex4-1}
\usepackage{amsmath,amscd,amsfonts,amssymb,graphicx}
\begin{document}
\title{Witnessing nonclassical correlations
via a single-shot experiment on an ensemble of spins
using NMR}
\author{Amandeep Singh}
\email{amandeepsingh@iisermohali.ac.in}
\affiliation{Department of Physical Sciences, Indian
Institute of Science Education and
Research Mohali, Sector 81 SAS Nagar, 
Manauli PO 140306 Punjab India.}
\author{Arvind}
\email{arvind@iisermohali.ac.in}
\affiliation{Department of Physical Sciences, Indian
Institute of Science Education  and
Research Mohali, Sector 81 SAS Nagar, 
Manauli PO 140306 Punjab India.}
\author{Kavita Dorai}
\email{kavita@iisermohali.ac.in}
\affiliation{Department of Physical Sciences, Indian
Institute of Science Education  and
Research Mohali, Sector 81 SAS Nagar, 
Manauli PO 140306 Punjab India.}
\begin{abstract}
A bipartite quantum system in a mixed state can exhibit
nonclassical correlations, which can go beyond quantum
entanglement.  While quantum discord is the standard
measure of quantifying such general quantum correlations,
the nonclassicality can be determined by simpler means via
the measurement of witness operators.  We experimentally
construct a positive map to witness nonclassicality of two
qubits in an NMR system.  The map can be decomposed so that
a single run of an experiment on an ensemble of spins
suffices to detect the nonclassicality in the state, if
present.  We let the state evolve in time and use the map to
detect nonclassicality as a function of time.  To evaluate
the efficacy of the witness operator as a means to detect
nonclassicality, we  measure quantum discord by performing
full quantum state tomography at each time point and
obtained a fairly good match between the two methods.
\end{abstract}
\maketitle
\section{Introduction}
Quantum correlations are those correlations which are not
present in classical systems, and in bipartite quantum
systems are associated with the presence of quantum
discord~\cite{nielsen-book-02,ollivier-prl-02,modi-rmp-12}.
Such correlations in a bipartite mixed state can go beyond
quantum entanglement and therefore can be present even if
the state is separable~\cite{ferraro-pra-10}.  The threshold
between classical and quantum correlations was investigated
in linear-optical systems by observing the emergence of
quantum discord~\cite{karo-pra-13}.  Quantum discord was
experimentally measured in systems such as NMR, that are
described by a deviation density
matrix~\cite{pinto-pra-10,maziero-bjp-13,passante-pra-11}.
Further, environment-induced sudden transitions in quantum
discord dynamics and their preservation were investigated
using NMR~\cite{auccaise-prl-11-2,harpreet-discord}.

It has been shown that even with very low (or no)
entanglement, quantum information processing  can still be
performed using nonclassical correlations as characterized
by the presence of quantum
discord~\cite{datta-pra-05,fahmy-pra-08}.  However,
computing and measuring quantum discord typically involves
complicated numerical optimization and 
furthermore it has
been shown that computing quantum discord is 
NP-hard~\cite{huang-njp-14}.
It is hence of prime
interest to find other means such as witnesses to detect
the presence of quantum correlations~\cite{saitoh-qip-11}.
While there have been several experimental implementations
of entanglement
witnesses~\cite{rahimi-jpamg-06,rahimi-pra-07,filgueiras-qip-11}, there
have been fewer proposals to witness nonclassicality.
A nonlinear classicality witness was constructed for a class
of two-qubit systems~\cite{maziero-ijqi-12} and
experimentally implemented using
NMR~\cite{auccaise-prl-11,pinto-ptrsa-12} and was estimated
in a linear optics system via statistics from a single
measurement~\cite{aguilar-prl-12}.  It is to be noted that
as the state space for classical correlated systems is not
convex, a witness for nonclassicality is more complicated to
construct than a witness for entanglement and is necessarily
nonlinear~\cite{saitoh-pra-08}.

In this work we report the experimental detection of
nonclassicality through a recently proposed positive map
method~\cite{rahimi-pra-10}.  The map is able to witness
nonclassical correlations going beyond entanglement, in a
mixed state of a bipartite quantum system. The method
requires much less experimental resources as compared to
measurement of discord using full state tomography and
therefore is an attractive alternative to demonstrating the
nonclassicality of a separable state.  The map
implementation involves two-qubit gates and single-qubit
polarization measurements and can be achieved in a single
experimental run using NMR.  We perform experiments on a
two-qubit separable state (non-entangled) which contains
nonclassical correlations.  Further, the state was allowed
to freely evolve in time under natural NMR decohering
channels, and the amount of nonclassicality present was
evaluated at each time point by calculating the map value.
We compared our results using the positive map witness with
those obtained by computing the quantum discord via full
state tomography, and obtained a good match.  However beyond
a certain time, the map was not able to detect
nonclassicality, although the quantum discord measure
indicated that nonclassicality was present in the state.
This indicates that while the positive map nonclassicality
witness is easy to implement experimentally in a single
experiment and is a good indicator of nonclassicality in a
separable state, it is not able to characterize
nonclassicality completely.  In our case this is typified by
the presence of a small amount of quantum discord when the
state has almost decohered or when the amount of
nonclassicality present is small.

The material in this paper is organized as
follows:~Section~\ref{mapvalue} contains a brief description
of the construction of the positive map to detect
nonclassicality, followed by details of the experimental NMR
implementation in Section~\ref{expt}.  The map value
dynamics with time is contained in Section~\ref{mv-dyn},
while the comparison of the results of the positive map
method with those obtained by measuring quantum discord via
full quantum state tomography is described in
Section~\ref{qd-dyn}.  Section~\ref{concl} contains some
concluding remarks.
\section{Experimentally detecting nonclassical correlations}
\label{fullexptl}
\subsection{Constructing the nonclassicality witness map}
\label{mapvalue}
For pure quantum states of a bipartite quantum
system which are represented by one-dimensional projectors
$\vert \psi\rangle\langle \psi\vert$ in a tensor
projector Hilbert space ${\cal H}_A \otimes {\cal
H}_B$, the only type of quantum correlation is
entanglement~\cite{guhne-pr-09,oppenheim-prl-02}.
However, for mixed states the situation is more
complex and quantum correlations can be
present even if the state is separable {\it i.\,e.}
it is a classical mixture of separable pure states
given by
\begin{equation}
\rho_{sep}=\sum_i{w_i\rho_i^A\otimes\rho_i^B} \label{sep}
\end{equation} 
where $w_i$ are positive weights and $\rho_i^A,\rho_i^B$ are
pure states in Hilbert spaces ${\cal
H}_A$ and  ${\cal H}_B$ respectively~\cite{peres-prl-96}.  
A separable state  is
called a properly classically correlated state
(PCC) if it can be written in the
form~\cite{horodecki-rmp-09} 
\begin{equation}
\rho_{\rm PCC}=\sum_{i,j}{p_{ij}\vert
e_i\rangle^{A}\langle e_i\vert\otimes\vert
e_j\rangle^{B}\langle e_j\vert} 
\label{pccform}
\end{equation} 
where $p_{ij}$ is a joint probability distribution
and $\vert e_i\rangle^{A}$ and $\vert
e_j\rangle^{B}$ are local orthogonal eigenbases in
local spaces ${\cal H}_A$ and  ${\cal H}_B$
respectively.  A state that cannot be written in
the form given by Equation~(\ref{pccform}) is
called a nonclassically correlated (NCC) state. An
NCC state can be entangled or separable.  The
correlations in NCC states over and above those in
PCC states are due to the fact that the eigenbases
for the subsystems may not be orthogonal {\it
i.\,e.} the basis vectors are in a
superposition~\cite{vedral-found-10}.  A typical
example of a bipartite two-qubit NCC state is
\begin{equation}
\sigma=\frac{1}{2}\left[\vert00\rangle\langle
00\vert+\vert1+\rangle\langle1+\vert\right]
\label{sigma} 
\end{equation} 
with $\vert + \rangle
= \frac{1}{\sqrt{2}}
\left(\vert 0 \rangle+\vert 1
\rangle\right)$.  In this case the state has
no product eigenbasis as the eigenbasis for
subsystem B, since $\vert 0 \rangle$ and
$\vert+\rangle$ are not orthogonal to each other.
The state is separable (not entangled) as it can
be written in the form given by
Equation~(\ref{sep}); however since it is an NCC 
state, it
has non-trivial quantum correlations and has
non-zero  quantum discord. How to pin down the
nonclassical nature of such a state with minimal
experimental effort and without actually computing
quantum discord is something that is desirable.
It has been shown that such  nonclassicality
witnesses can be constructed using a positive
map~\cite{rahimi-pra-10}.

The map $\mathcal{W}$ over the state
space ${\cal H}={\cal H}_A \otimes \cal{H}_B$ 
takes a state 
to a real number $\mathcal{R}$ 
\begin{equation}
\mathcal{W}:{\cal H}\longrightarrow\mathcal{R} 
\label{map}
\end{equation} 
This map is a nonclassicality witness map {\it i.\,e.} it is
capable of detecting NCC states in ${\cal H}$
state space if and only if~\cite{rahimi-pra-10}: 
\begin{itemize}
\item[(a)] For every bipartite
state $\rho_{PCC}$ having a product eigenbasis,
$\mathcal{W}:\rho_{PCC}\geq0$. 
\item[(b)] There exists
at least one bipartite state $\rho_{NCC}$ (having no
product eigenbasis) such that $\mathcal{W}:\rho_{NCC}<0$.
\end{itemize}
A specific non-linear nonclassicality witness map proposed
by~\cite{rahimi-pra-10} 
is defined in terms of expectation
values of positive Hermitian operators  $\hat{A}_1$,
$\hat{A}_2\ldots\hat{A}_m$:
\begin{equation}
\mathcal{W} : \rho\rightarrow c-\left(Tr(\rho\hat{A}_1)\right)
\left(Tr(\rho\hat{A}_2)\right)\ldots\ldots\left(Tr(\rho\hat{A}_m)\right)
\end{equation} 
where $c \ge 0$ is a real number.

For the case of two-qubit systems
using the operators $A_1=\vert00\rangle\langle
00\vert$ and $A_2=\vert 1+\rangle\langle
1+\vert$ we obtain a nonclassicality witness map
for state in Eqn.~(\ref{sigma}) as:
\begin{equation}
\mathcal{W}_{\sigma}:\rho\rightarrow c-\left(Tr(\rho\vert00\rangle\langle
00\vert)\right)\left(Tr(\rho\vert1+\rangle\langle
1+\vert)\right) \label{sigmamap} 
\end{equation} 
The value of the constant $c$ in the above witness
map has to be optimized  such that for any PCC state
$\rho$ having a product eigenbasis, the condition
$\mathcal{W}_{\sigma} \rho \ge 0$ holds and the
optimized value of  $c$ turns out to be
$c_{\rm opt}=0.182138$.
The map given by Equation~(\ref{sigmamap}) does
indeed witness the nonclassical nature of the state $\sigma$ as
$\left(Tr(\rho\vert00\rangle\langle
00\vert)\right)\left(Tr(\rho\vert 1+\rangle\langle
1+\vert)\right)$ for $\rho\equiv\sigma$ has the
value 0.25, which suggests that the state
$\sigma$ is an NCC state~\cite{rahimi-pra-10}. 
The value of a nonclassicality map which when negative
implicates the nonclassical nature of the state is 
denoted its map value (MV). 
\subsection{NMR experimental system}
\label{expt}
\begin{figure}[t]
\includegraphics[angle=0,scale=1.0]{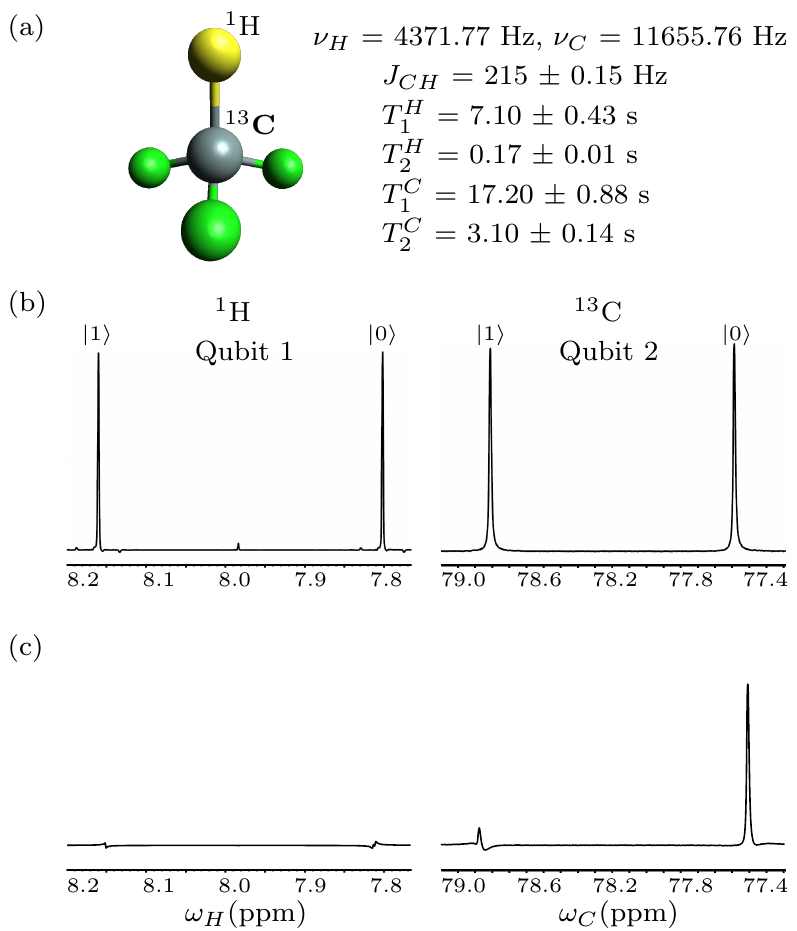}
\caption{(a) Pictorial representation of ${}^{13}$C
labeled chloroform with the two qubits encoded as nuclear
spins of ${}^{1}$H and ${}^{13}$C; system
parameters including chemical shifts $\nu_i$, scalar coupling 
strength $J$
(in Hz) and relaxation times T$_{1}$ and T$_{2}$ (in
seconds) are tabulated alongside. 
(b) Thermal equilibrium NMR spectra of ${}^{1}$H
(Qubit 1) and ${}^{13}$C (Qubit 2) after a $\frac{\pi}{2}$
readout pulse. (c) NMR spectra of  ${}^{1}$H and ${}^{13}$C for
the $\sigma$ NCC state. Each transition in the spectra is labeled 
with the logical
state ($\vert0\rangle$ or $\vert 1\rangle$) of the ``passive
qubit'' (not undergoing any transition).} 
\label{molecule}
\end{figure}

We implemented the nonclassicality witness map
$\mathcal{W}_\sigma$   on
an NMR sample of ${}^{13}$C-enriched
chloroform dissolved in acetone-D6; the ${}^{1}$H
and ${}^{13}$C nuclear spins were used to encode
the two qubits (see Fig.~\ref{molecule} for experimental
parameters). 
Unitary operations were implemented by specially crafted
transverse radio frequency pulses of suitable amplitude,
phase and duration.  
A sequence of spin-selective
pulses interspersed with tailored free evolution periods were
used to prepare the system in an NCC state as
follows:
\begin{eqnarray*}
&&I_{1z}+I_{2z}
\stackrel{(\pi/2)^1_x}{\longrightarrow}
-I_{1y}+I_{2z}
\stackrel{Sp. Av.}{\longrightarrow}
I_{2z}
\stackrel{(\pi/2)^2_y}{\longrightarrow}
I_{2x} 
\stackrel{\frac{1}{4J}}{\longrightarrow}
\nonumber\\
&&
\quad\quad\quad\quad
\frac{I_{2x}+2I_{1z}I_{2y}}{\sqrt{2}}
\stackrel{(\pi/2)^2_x}{\longrightarrow}
\frac{I_{2x}+2I_{1z}I_{2z}}{\sqrt{2}}
\stackrel{(-\pi/4)^2_y}{\longrightarrow}
\nonumber \\
&&
\quad\quad\quad\quad
\quad\quad\quad
\quad
\frac{\left(I_{2z}+I_{2x}+2I_{1z}I_{2z}-2I_{1z}I_{2x}\right)}{2}
\end{eqnarray*} 
\begin{figure}[h]
\includegraphics[angle=0,scale=1]{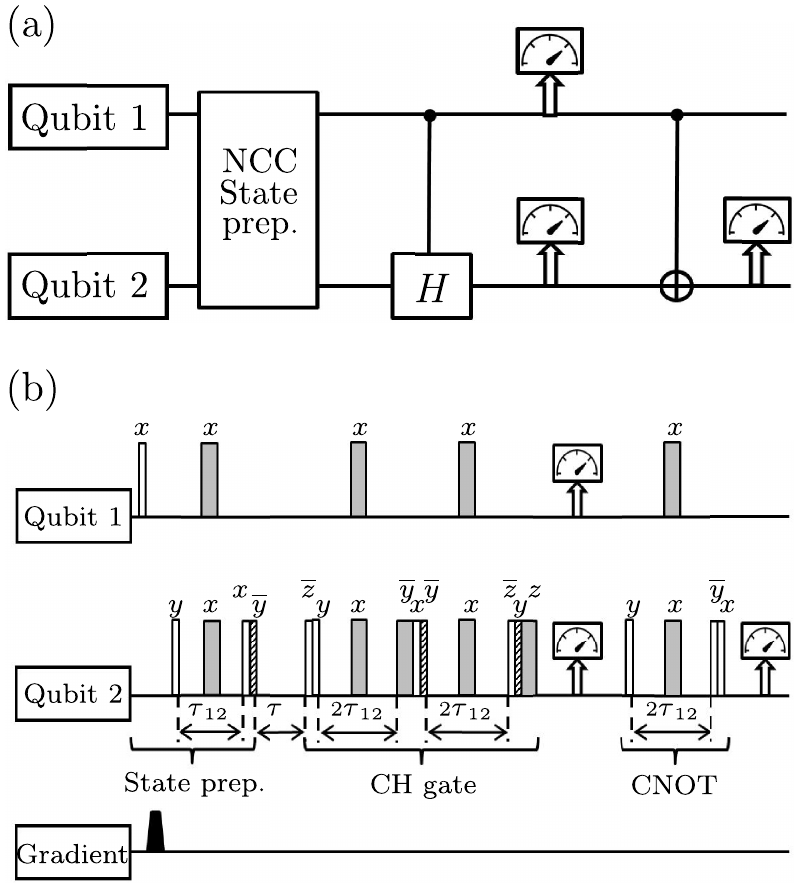} 
\caption{(a)
Quantum circuit to create and detect an NCC state. 
(b) NMR
pulse sequence to create the NCC state and then
detect it using controlled-Hadamard and CNOT gates. Unfilled
rectangles depict $\frac{\pi}{2}$ pulses, grey-shaded rectangles 
depict $\pi$ pulses and filled rectangles depict $\frac{\pi}{4}$
pulses, respectively. Phases are written above each pulse,
with a bar over
a phase indicating a negative phase. 
The evolution period was set to $\tau_{12}=\frac{1}{4J}$
(refer main text for details of delay $\tau$).} 
\label{ckt} 
\end{figure}
The quantum circuit to implement the nonclassicality witness
map is shown in Fig.~\ref{ckt}(a).  The first module
represents NCC state preparation using the pulses as already
described. The circuit to capture nonclassicality of the
prepared state consists of a controlled-Hadamard (CH)  gate,
followed by measurement on both qubits, a controlled-NOT
(CNOT) gate and finally detection on `Qubit 2'. The CH gate
is analogous to a CNOT gate, with a Hadamard gate being
implemented on the target qubit if the control qubit is in
the state $\vert 1 \rangle$ and a `no-operation' if the
control qubit is in the state $\vert 0 \rangle$.  The NMR
pulse sequence corresponding to the quantum circuit is
depicted in Fig.~\ref{ckt}(b). The set of pulses grouped
under the label `State prep.' convert the thermal
equilibrium state to the desired NCC state. A dephasing
$z$-gradient is applied on the gradient channel to kill
undesired coherences.  After a delay $\tau$ followed by the
pulse sequence to implement the CH gate, the magnetizations
of both qubits were measured with $\frac{\pi}{2}$ readout
pulses (not shown in the figure).  In the last part of
detection circuit a CNOT gate is applied followed by a
magnetization measurement of `Qubit 2'; the scalar coupling
time interval was set to $\tau_{12}=\frac{1}{4J}$ where $J$
is the strength of the scalar coupling between the qubits.
Refocusing pulses were used during all $J$-evolution to
compensate for unwanted chemical shift evolution during the
selective pulses.  State fidelity was computed using the
Uhlmann-Jozsa measure~\cite{uhlmann-rpmp-76,jozsa-jmo-94},
and the NCC state was prepared with a fidelity of $0.97
\pm 0.02$.

To detect the nonclassicality in the prepared NCC
state via the map $\mathcal{W}_\sigma$, the
expectation values of the operators
$\vert00\rangle\langle 00\vert$ and $\vert
1+\rangle\langle 1+\vert$ are required.  Re-working
the map brings it to the following
form~\cite{rahimi-pra-10}
\begin{eqnarray*}
\mathcal{W}_{\sigma}:\rho\rightarrow 
c_{\rm opt}-\frac{1}{16}&&\left(1+\langle
Z_1\rangle+\langle Z_2\rangle+\langle
Z_2'\rangle\right)\times \nonumber \\
&&\left(1-\langle Z_1\rangle+\langle Z_2\rangle-\langle
Z_2'\rangle\right) 
\end{eqnarray*}
where $\langle Z_1\rangle$ and $\langle
Z_2\rangle$ are the polarizations of `Qubit 1' and
`Qubit 2' after a CH gate on the input state
$\rho$, while $\langle Z_2'\rangle$ is the
polarization of `Qubit 2' after a CNOT gate.  The
theoretically expected normalized values of
$\langle Z_1\rangle$, $\langle Z_2\rangle$ and
$\langle Z_2'\rangle$ for state $\rho\equiv\sigma$
are $0$, $1$ and $0$ respectively. 
Map value (MV) is $-0.067862<0$ and as desired this map does
indeed witness the presence of nonclassicality.
The experimentally computed MV for the prepared
NCC state turns out to be $-0.0406 \pm 0.0056$,
proving that the map is indeed able to witness the
nonclassicality present in the state.
\subsection{Map Value Dynamics} 
\label{mv-dyn}
\begin{figure}[h] 
\includegraphics[angle=0,scale=1]{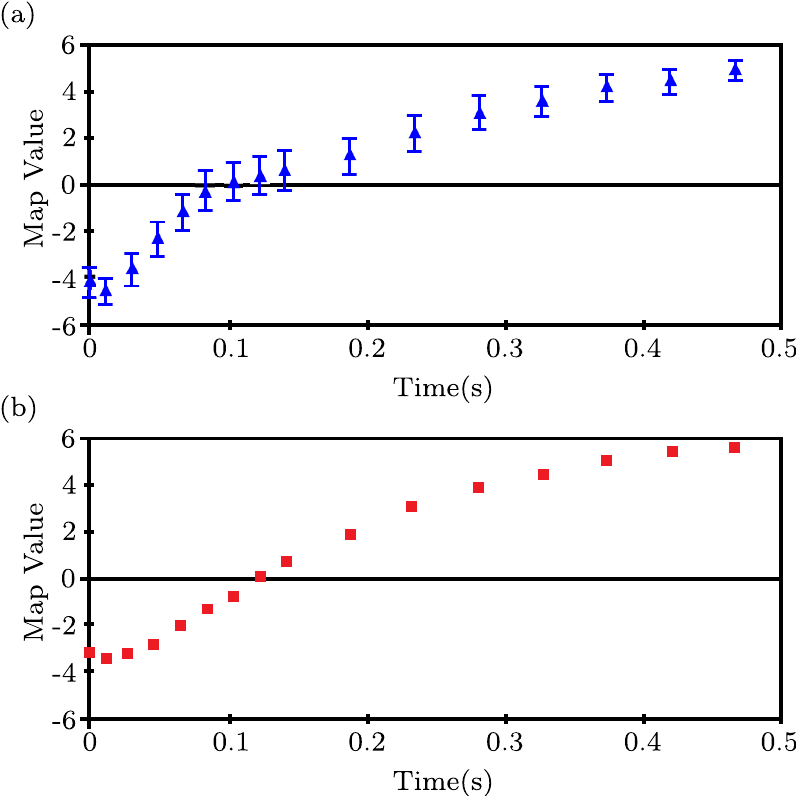}
\caption{(a) Experimental map value (in
$\times10^{-2}$ units) plotted as a function of time. (b) Map value
(in $\times10^{-2}$ units) directly calculated from
the tomographically reconstructed state at each time point.}
\label{MV} 
\end{figure}

\begin{figure}[h]
\includegraphics[angle=0,scale=1]{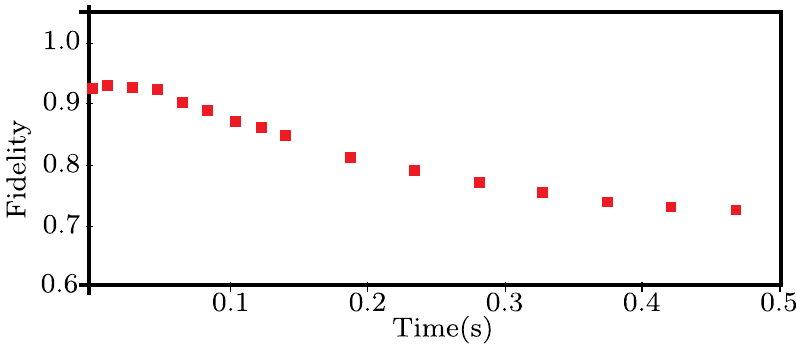} 
\caption{(Color online)
Time evolution of state fidelity. The red squares
represent fidelity of the experimentally prepared
NCC state w.r.t. the
the theoretical NCC state.}
\label{fid} 
\end{figure}

The prepared NCC state was allowed to evolve freely in time
and the MV calculated at each time point,  in order to
characterize the decoherence dynamics of the nonclassicality
witness map. As theoretically expected, one should get a
negative MV for states which are NCC.  We measured MV at
time points which were integral multiples of $\frac{2}{J}$
i.e. $\frac{2n}{J}$ (with $n$ = 0, 1, 3, 5, 7, 9, 11, 13,
15, 20, 25, 30, 35, 40, 45 and 50), in order to avoid
experimental errors due to $J$-evolution. The results of MV
dynamics as a function of time are shown in Fig.~\ref{MV}(a).  
The standard NMR decoherence mechanisms
denoted by T$_2$ the spin-spin relaxation time and T$_1$ the
spin-lattice relaxation time, cause dephasing among the
energy eigenstates and energy exchange between the spins and
their environment, respectively.
As seen from Fig.~\ref{MV}(a), 
the MV remains negative (indicating the state is NCC) for
upto 120 ms, which is approximately the ${}^{1}$H transverse
relaxation time.
The MV was also
calculated directly from the tomographically reconstructed
state using full state
tomography~\cite{leskowitz-pra-04} at each time point and 
the results are shown in Fig.~\ref{MV}(b), which are
in good agreement with direct experimental MV measurements.
The state
fidelity was also computed at the
different time points and the results 
are shown in Fig.~\ref{fid}.
The red squares represent fidelity w.r.t.
the theoretical NCC state.
\subsection{Quantum Discord Dynamics} 
\label{qd-dyn}
We also compared the map value evaluation of nonclassicality
with the standard measure of nonclassicality, namely
quantum discord~\cite{ollivier-prl-02,luo-pra-08}.
The state was reconstructed by performing full
quantum state tomography and the 
quantum discord measure was computed from the
experimental data.
Quantum mutual information can be quantified by the
equations:
\begin{eqnarray}
I(\rho_{AB})&=&S(\rho_A)+S(\rho_B)-S(\rho_{AB})
\nonumber \\
J_A(\rho_{AB})&=&S(\rho_B)-S(\rho_B\vert \rho_A)
\label{mutual}
\end{eqnarray}
where $S(\rho_B\vert\rho_A)$ is the conditional von Neumann
entropy of subsystem $B$ when $A$ has already 
been measured.
Quantum discord is defined as the minimum difference between
the two formulations of mutual information in
Equation~(\ref{mutual}):
\begin{equation}
D_A(\rho_{AB})=S(\rho_A)-S(\rho_{AB})+S(\rho_B\vert\lbrace\Pi^A_j\rbrace)
\label{discord} 
\end{equation} 
Quantum discord hence depends on projectors
$\lbrace\Pi^A_j\rbrace$. 
The
state of the system, after the outcome corresponding to
projector $\lbrace\Pi^A_j\rbrace$ has been detected, is
\begin{equation} \tilde{\rho}_{AB}\vert
\lbrace\Pi^A_j\rbrace=\frac{\left(\Pi^A_j\otimes
I_B\right)\rho_{AB}\left(\Pi^A_j\otimes I_B\right)}{p_j}
\label{B} \end{equation} with the probability
$p_j=Tr\left((\Pi^A_j\otimes I_B)\rho_{AB}(\Pi^A_j\otimes
I_B)\right)$; $I_B$ is identity operator on subsystem B.
The state of the system B, after this measurement is
\begin{equation} \rho_B\vert
\lbrace\Pi^A_j\rbrace=Tr_A\left(\tilde{\rho}_{AB}\vert
\lbrace\Pi^A_j\rbrace\right) \label{A} \end{equation}
$S\left(\rho_B\vert\lbrace\Pi^A_j\rbrace\right)$ is the
missing information about B before measurement
$\lbrace\Pi^A_j\rbrace$. The expression \begin{equation}
S(\rho_B\vert\lbrace\Pi^A_j\rbrace)=\sum_{j}{p_jS
\left(\rho_B\vert\lbrace\Pi^A_j\rbrace\right)}
\label{cond-entropy} 
\end{equation} 
is the conditional entropy
appearing in Eqn.~(\ref{discord}).  
In order to capture the true quantumness of the correlation
one needs to perform an optimization over all sets of von
Neumann type measurements represented by the projectors
$\lbrace\Pi^A_j\rbrace$. 
We define two orthogonal vectors (for spin half quantum
subsystems), characterized by two real parameters $\theta$
and $\phi$, on the Bloch sphere as:
\begin{eqnarray}
&&\cos{\theta}\vert 0\rangle+e^{\iota\phi}\sin{\theta}\vert
1\rangle  \nonumber \\
&&e^{-\iota\phi}\sin{\theta}\vert 0\rangle-\cos{\theta}\vert
1\rangle 
\end{eqnarray} 
These vectors can be used to
construct the projectors $\Pi^{A,B}_{1,2}$,
which are then used to find out the state of
$B$ after an arbitrary measurement was made on
subsystem $A$. The definition of conditional
entropy (Equation~(\ref{cond-entropy})) can be used
to obtain an expression which is parameterized by $\theta$
and $\phi$ for a given state $\rho_{AB}$. This expression is
finally minimized by varying $\theta$ and $\phi$ and the
results fed back into Equation~(\ref{discord}), which yields
a measure of quantum discord independent of the basis chosen
for the measurement of the subsystem.

\begin{figure}[h] \includegraphics[angle=0,scale=1]{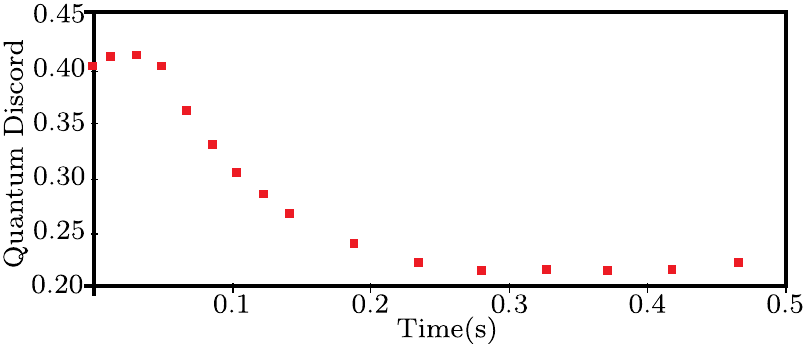}
\caption{(Color online) Time evolution of quantum discord 
(characterizing total
quantum correlations present in the state) for 
the NCC state.}
\label{QD} 
\end{figure}

To compare the detection via the positive map method 
with the standard quantum discord measure, we
let the state evolve for a time $\tau$ and then
reconstructed
the experimentally prepared 
via full quantum state tomography and calculated the
quantum discord at all time points
where the MV was
determined experimentally (the   
results are shown in Fig.~\ref{QD}).
At $\tau$ = 0 s, a non-zero QD confirms the presence of NCC
and verifies the results given by MV. As the
state evolves with time, the quantum
discord parameter starts decreasing rapidly,
in accordance with increasing MV. Beyond 120 ms, while the
MV becomes positive and hence fails to detect
nonclassicality, the discord parameter remains non-zero,
indicating the presence of some amount of nonclassicality (although by
this time the state fidelity has decreased to 0.7). However,
value of quantum discord is very close to zero and in fact
cannot be distinguished from contributions due to noise.
One can hence conclude that the positive map suffices to
detect nonclassicality when decoherence processes have not
set in and while the fidelity of the prepared state is good.
Once the state has decohered however, a measure such as 
quantum discord has to be used to verify if the degraded
state retains some amount of nonclassical correlations or
not.
\section{Conclusions}
\label{concl}
In this work we experimentally detected nonclassical
correlations in a separable two-qubit quantum state, using a
nonlinear positive map as a nonclassicality witness.  The
witness is able to detect nonclassicality in a single-shot
experiment and its obvious advantage lies in its using much
fewer experimental resources as compared to quantifying
nonclassicality by measuring discord via full quantum state
tomography.  It will be interesting to construct and utilize
this map in higher-dimensional quantum systems and for
greater than two qubits, where it is more difficult to
distinguish between classical and quantum correlations.

It has been posited that quantum correlations captured by
quantum discord which go quantum entanglement and can thus
be present even in separable states are responsible for
achieving computational speedup in quantum algorithms. It is
hence important, from the point of view of quantum
information processing, to confirm the presence of such
correlations in a quantum state, without having to expend
too much experimental effort and our work is a step forward
in this direction.

\begin{acknowledgments}
All experiments were performed on a Bruker Avance-III 600
MHz FT-NMR spectrometer at the NMR Research Facility at
IISER Mohali. Arvind acknowledges funding from DST India
under Grant No.~EMR/2014/000297. KD acknowledges funding
from DST India under Grant No.~EMR/2015/000556.
\end{acknowledgments}

%
\end{document}